# Three sources and three components of success in detection of ultra-rare alpha decays at the Dubna Gas-Filled Recoil separator

Yu.S.Tsyganov[1]

*FLNR, JINR, 141980 Dubna, Russia*

*tyura@sungns.jinr.ru*

**Abstract**

*General philosophy of procedure of detecting rare events in the recent experiments with $^{48}$Ca projectile at the Dubna Gas-Filled Recoil Separator(DGFRS) aimed to the synthesis of superheavy elements (SHE) has been reviewed. Specific instruments and methods are under consideration. Some historical sources of the successful experiments for Z=112-118 are considered too. Special attention is paid to application of method of "active correlations" in heavy-ion induced complete fusion nuclear reactions. Example of application in Z=115 experiment is presented. Brief description of the $^{243}$Am + $^{48}$Ca $\rightarrow$ $^{291-x}$115+xn experiment is presented too. Some attention is paid to the role of chemical experiments in discoveries of SHEs. The DGFRS detection/monitoring system is presented in full firstly.*

## 1. Introduction

In experiments [1-3] performed at the Flerov Laboratory of Nuclear Reactions, JINR, 45 new isotopes of new chemical elements with the atomic number 112-118 have been synthesized. The complete fusion reactions resulting from accelerated $^{48}$Ca ions bombarding the targets of $^{238,233}$U, $^{242,244}$Pu, $^{243}$Am, $^{245,248}$Cm, $^{237}$Np, $^{249}$Bk and $^{249}$Cf were used for the production of the above superheavy elements (SHEs). The nuclei of new elements appeared to mainly undergo α-decays (one or several) until the decay chain ends with a spontaneous fission (SF). Thin targets layers (~0.3-0.4 mg/cm$^2$) of highly enriched actinide isotopes were irradiated with $^{48}$Ca ion beams of a strictly preset energy. Recoil nuclei knocked out of the target were separated in flight from the $^{48}$Ca ions and various reaction products by means of the Dubna Gas-Filled Recoil Separator (DGFRS) [4], which was tuned to transmit the complete fusion products with an efficiency of about 40%. Note, that for exact tuning of the separator the equilibrium charge states systematic for low pressure hydrogen (one Torr) [5] was used. To separate the separator working volume with respect to the U 400 cyclotron high vacuum (~$10^{-6}$ to $10^{-7}$ Torr) the rotating entrance 1.5μm Ti-foil window has been used [6].
Typical intensity of the ion beam was about 0.8-1.5 pμA at the actinide target.

---

[1] *tyura@sungns.jinr.ru*

The mostly significant role in the detection process played the integrated detection-parameter monitoring/protection system of the DGFRS [7-11]. Namely with this system it has became possible to detect ultra-rare decays of SHEs.

## 2. α-α correlation chains: brief historical overview

The decays of nuclei were registered by position-sensitive PIPS detector mounted in the separator focal plane. Correlated decays of single atoms, i.e., chains of sequential α-decays terminated by a spontaneous fission event (α-α-α…SF), registered by the detector array are interpreted as decay sequences of unknown nuclides. Their identification is based on their radioactive decay properties and the reaction mechanism, in particular, on the characteristic dependence of the yield of neutron-evaporation products on the excitation energy of the compound nucleus. Investigating these dependences requires time-consuming measurements of the production cross-sections of the nuclei of interest at various energies of $^{48}$Ca ion beam. It was D.Parfanovich who first recognized the role of α-α… correlation chains to this aim.

He epitomized a principle for Z=102 element identification in the following form:
"There exist extra opportunities of physical identification of the new nuclei. In particular, registering the decay of element 102 and its daughter element 100 will allow determining more reliably the atomic and mass number of the new element. The essence of the method is that in the photographic plate one selects the tracks of alpha-particles that originate from the same point and have the energy of 8 and 9 MeV. Once such a "fork" is found the identification becomes stricter because of the tracks correspond to the genetically linked alpha-decays of elements 102 and 100 [12]. Of course, at that time it was not possible to measure α - decay energy precisely and only energy-position (no energy-time-position) correlation was measured.

The significant contribution in the development of this method was performed by Hoffmann [13]. It was SHIP team who introduced into experimental practice the so called approach *"one event-one element"* that is using of very restricted statistics to measure and identify of new synthesized element or isotope. It was a revolution in the detection/discovery method.

The third source was a method developed by Kharitonov [14] that means using α-α correlation chains to suppress background signals, although in the "passive"[2] mode. When processing stored data in search for 6.75-MeV $^{246}$Cf - a product under investigation some signals were excluded on the basis of finding α-α correlated signals with time interval less or equal to 1 s thus eliminating background from $^{223}$Ra → $^{219}$Rn (6.82 MeV )→$^{215}$Po. This method of background suppression was effective only with detecting geometry close to 4π.

---
[2] It means, only when off-line data analysis has been performed.

### 3. Sum of technologies to detect ultra rare events at the DGFRS

Of course, to measure the first correlation time the time-of-flight (or/and ΔE) device operating together with the PIPS detector to detect ER (recoil) signal was strongly required. It was Mezentsev, who first develop *low pressure thin window pentane-filled* (~1.5 Torr) TOF device for the DGFRS [15]. It consists of two (START and STOP) multi - wire chambers for ER signal high efficiency detection. After application of this TOF module in test reactions, like $^{nat}$Yb + $^{48}$Ca → *Th, $^{197}$Au+$^{22}$Ne→ *Ac and others, it was applied in the long term experiments (month-months-half year an actinide target irradiation continuously) aimed to the synthesis of SHE. Below, in the Fig.1 result of one of the test is shown. Together with main histogram of alpha decay spectrum from $^{nat}$Yt + $^{48}$Ca → *Th nuclear reaction, in the top line the on-line measurement test of the $^{48}$Ca ions suppression factor is shown for statistics of N=729649 particles.

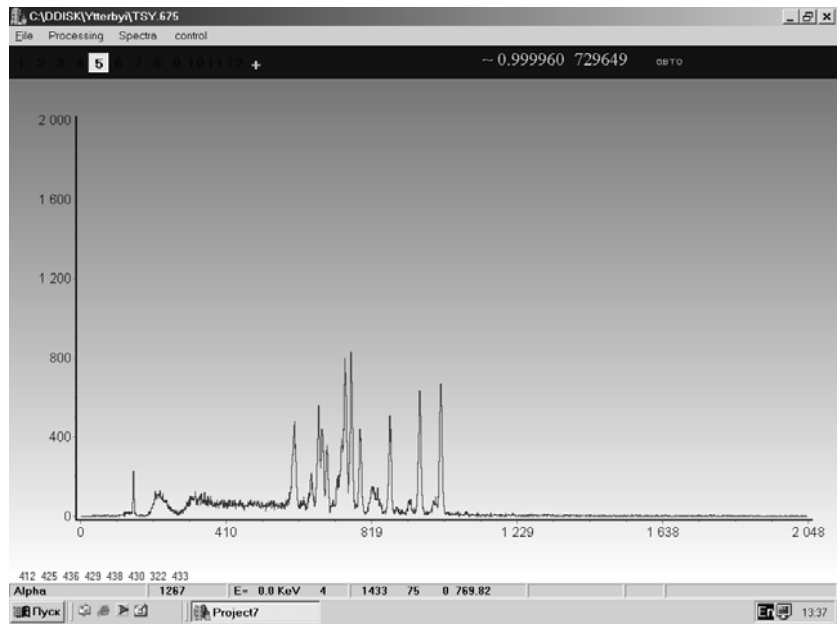

**Fig.1** Alpha decay spectrum (for strip#5 of PIPS detector-left-upper line) from $^{nat}$Yb+$^{48}$Ca →*Th nuclear reaction. Efficiency of $^{48}$Ca ions suppression value is shown in upper part (0.99996 for 729649 ions).

Very significant detection component is application of method of *"active correlations"* [16-21] which provided a radical background products associated with the cyclotron beam. It allowed us to employ a special low-background scheme for the nuclei to be investigated. The beam was switched off after a recoil signal was detected with parameters of implantation energy expected for Z=115 evaporation residues, followed by an α-like signal with an energy 9800 ≤ E$_α$

≤ 11000 , in the same strip, within ~2.2- mm-wide position. The triggering ER-α time was set according to formula $\Delta t = 0.25 \cdot (11000 - E_\alpha)$ [KeV]. The beam-off interval initially set at 1 min. In this time interval, if α particle with 8800≤Eα≤10800 KeV interval registered in any position of the same strip, the beam-off interval was automatically extended to 3 min. Namely with development and application of this method experimentalist firstly measured in background free mode the decaying spectra of SHE (Fig.'s 2-4 as example of application).

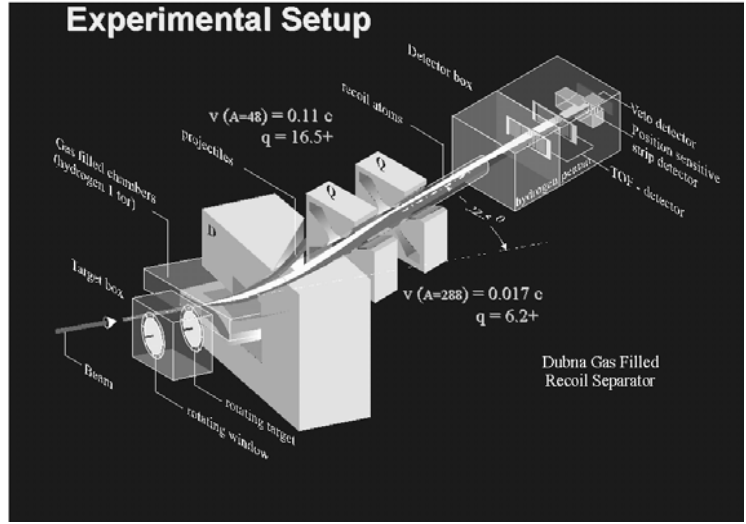

**Fig.2** The experimental setup - Dubna Gas-Filled Recoil Separator.

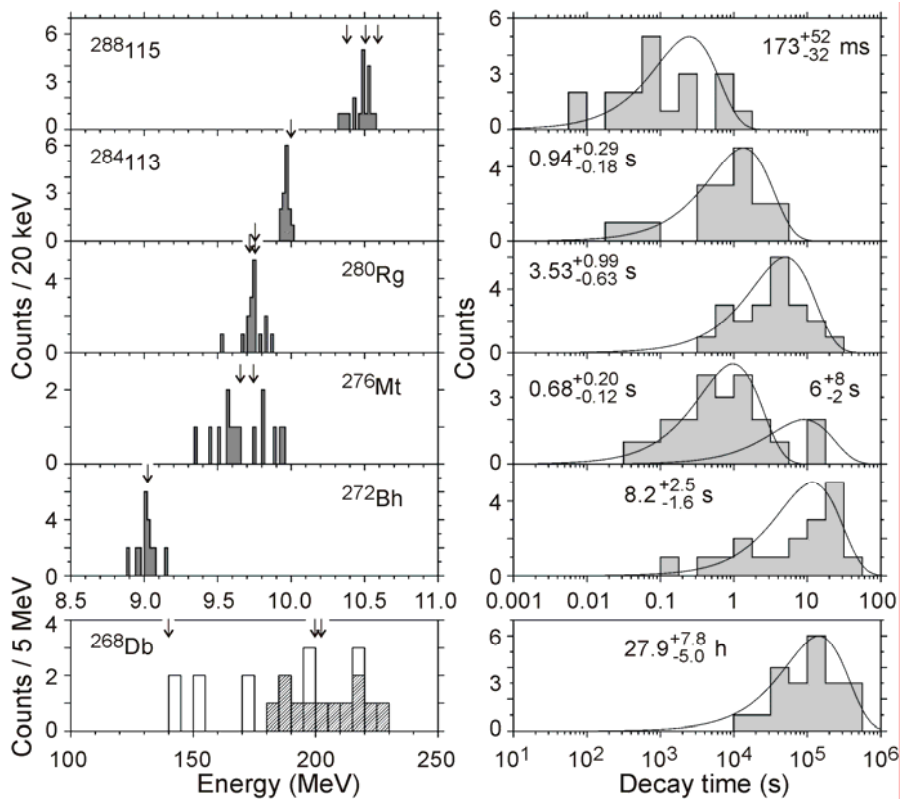

**Fig.3** Measured decay spectra for $^{243}$Am+$^{48}$Ca→ $^{288}$115 + 3n recent experiment [22].
Arrows indicate result of the previous experiment (2003). Dashed bins for $^{268}$Db SF spectrum correspond to detection of both fragments (by PIPS focal plane detector and by backward one).

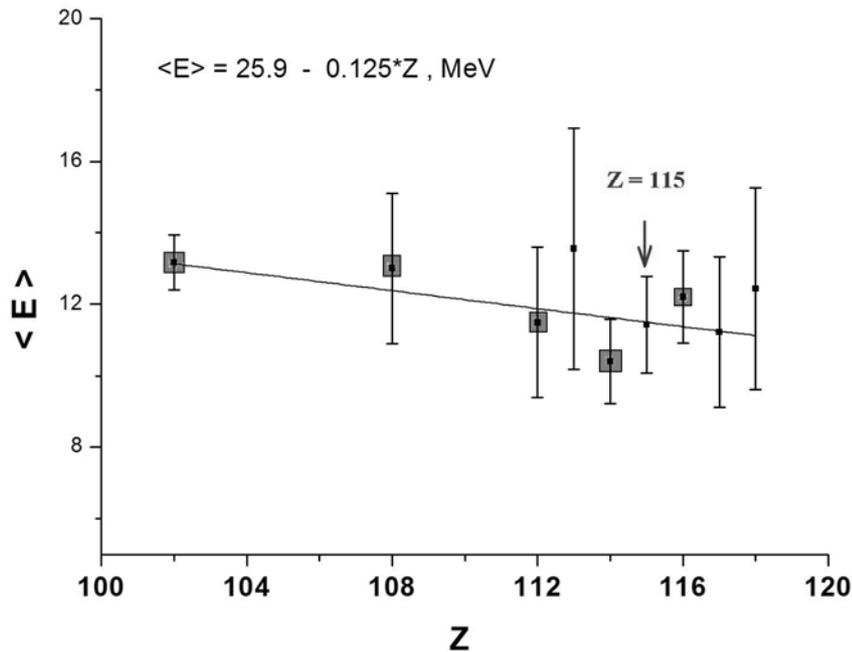

**Fig.4** Dependence of ER registered energy against Z for the DGFRS detecting module and for complete fusion reactions $^{48}$Ca + $^A$Target $\rightarrow$ $^{(A+48-x)}$ SHE +xn . Gray rectangles indicate known elements. Z=115 is shown by an arrow.

And finally, when performing a long-term experiment aimed at the detection of rare (maybe even single) decays of superheavy nuclei, one should definitely use not only a detection system, but an integrated detection-parameter monitoring-protection system to avoid unreasonable scenarios, especially associated with alarm situations. This requirement becomes the stronger as the more active actinide target is used in the experiment and more intense beams are applied.

In FLNR such a system was designed and applied since 2009 after more than two years of tests. This is this system that was used in synthesizing element Z=117 [23]. In the Fig. 5,6 a general schematics (**5**) of the PC-based system and main user interface (**6**) of the control subsystem [24] are shown. It consists of four subsystems, namely:

(1) *Detection system;*

(2) *Parameter monitoring and protection system;*

(3) *Cyclotron beam manipulation system;*

(4) *Visualization (remote control room);*

(5) *Target volume aerosol control system;*

(6) *Beam energy measurement system.*

-

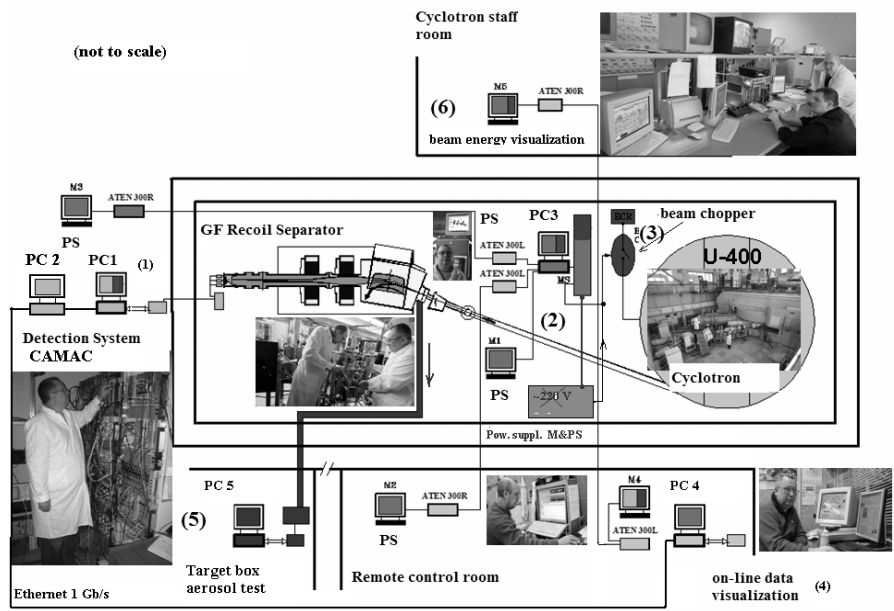

**Fig.5** Lay-out of integrated detection-monitoring and protection system of the DGFRS

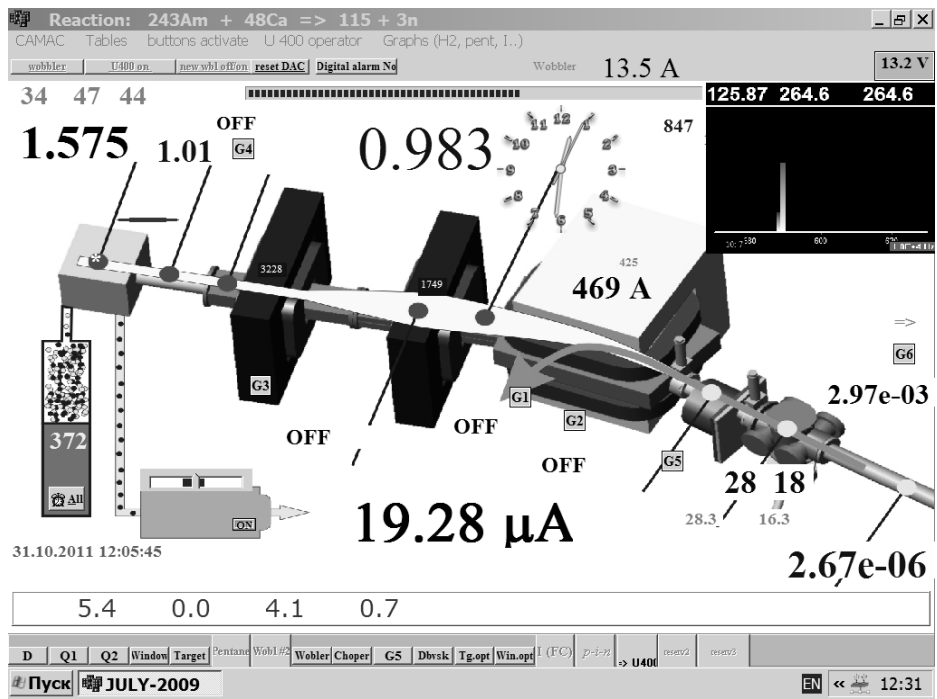

**Fig.6** The DGFRS control system interface. ( coded *in C++ Builder, Windows XP*). Right-upper window: $^{48}$Ca beam energy monitoring (the same histogram/digits exist at cyclotron staff room – refresh exposition time ~0.6 s).

In this figure current values of the following main parameters are shown:

19.28 µA- the DGFRS Faraday cup $^{48}$Ca ion beam intensity, 34 – rate of events ($c^{-1}$), 47/44 START/STOP TOF module signal rates, respectively, 28 $c^{-1}$ – rotation speed for actinide target wheel, 18 – the same one for rotating entrance window, 2.67e-06 – cyclotron vacuum ( Torr, ~2.5 m apart from the DGFRS entrance window), 1.575 Torr –pentane pressure in the TOF module, 1.01 and 0.983 – hydrogen pressure, 469 A – dipole magnet current value, 13.5 A – beam wobbling system winding current value, 372 (Torr)- pentane vapour pressure in the liquid pentane volume, 264.6 (MeV) – beam energy value, 5.1 (atm.) – water pressure in dipole magnet cooling system, 4.1 (atm.) – the same input value for rotating target cooling. The string " 31.10.2011 12:05:45 " indicates data/time of the last beam stop after searching for a pointer to ER-alpha correlation sequence.

Example of one of visual system windows is shown in the Fig.7. It shows what experimentalist could see a few tens of seconds after detection system registered a pointer to ER-α correlation in real-time mode.

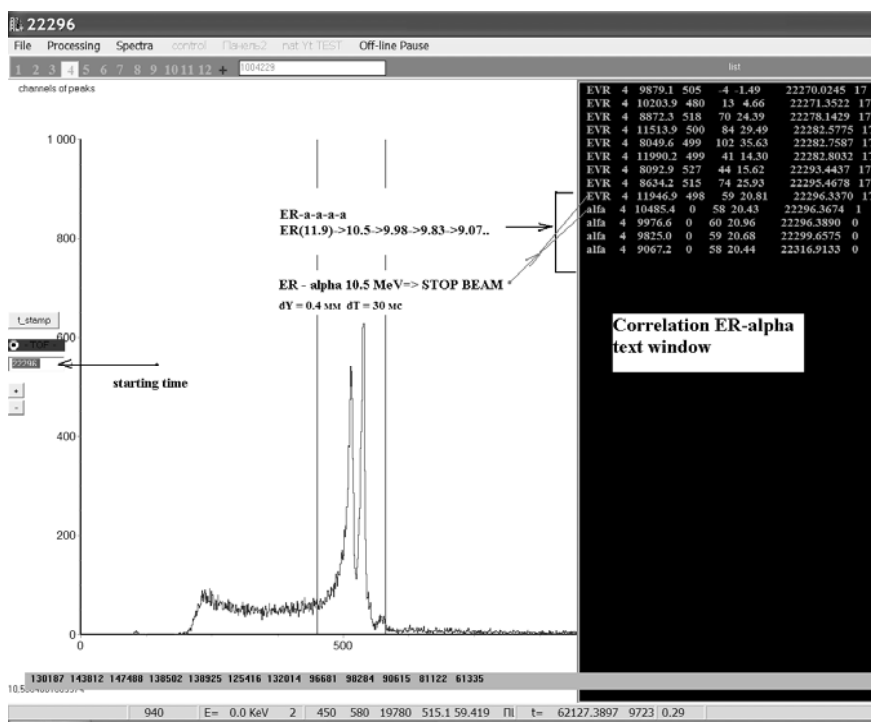

**Fig.7** Example of on-line/real-time search for ER-α energy - time –position correlation for Z=115. Marker "0" in the text window means Beam-OFF interval, the same value with "1" – Beam-ON. Right column denotes elapsed time in seconds.

## 4. On the role of chemistry experiments

Chemical identification of isotopes in the observed decay chains could give us the identity of the atomic numbers of nuclei in the decay chain and provide independent evidence for the discovery of a new element (isotope) [25,26]. An isotope of the element 115 with the mass number 288 undergoes five sequential α-decays 115→113→111→109→107→105 (SF) ending with the spontaneous fission of $^{268}$Db. The half-life of $^{268}$Db is high enough for the liquid chemistry procedure.

Another way is to study isotopes under researcher's interest using gaseous chemistry [27]. Note, last years significant progress was achieved in this way. As an example in the Fig.'s 8,9 two chains of Z=111 isotope were presented [28]. This multi-chain event was measured in the $^{243}$Am+$^{48}$Ca → $^{288}$115+3n complete fusion nuclear reaction. The main reason, why the event started from Z=111 and not from Z=115 is the relatively high dead time of the chemical transport procedure (1-2… seconds) compared with electromagnetic separator.

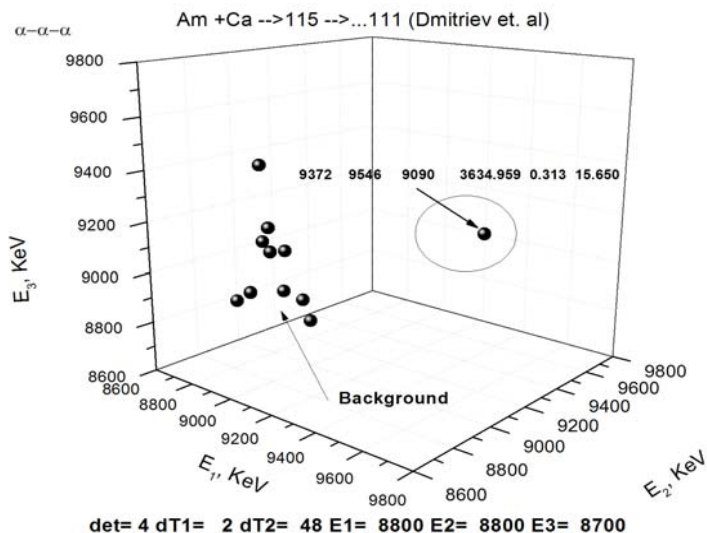

**Fig.8** 3D picture of α-α-α correlation measured in $^{243}$Am+$^{48}$Ca → $^{288}$115+3n reaction. Correlation time windows are: 4 s and 48 s for a first and second α-α chain, respectively. Measured correlation times are 0.313 and 15.65 s, respectively [28].

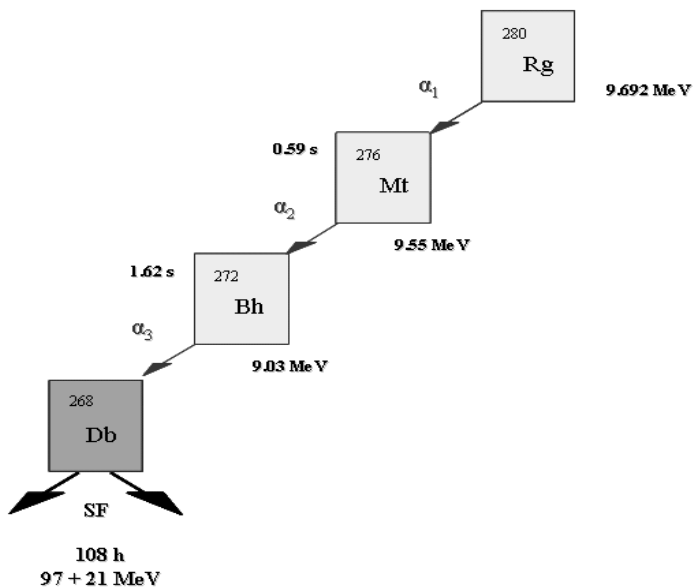

**Fig.9** The second event measured in chemical experiment [28].

Note, that a positive role of chemistry in the discoveries of SHE is in progress now. In the given case (Z=115) both chains indicate to the independent confirmation of the experimental results at the DGFRS.

**5. Summary**

Three main sources and components for detection of ultra rare alpha decays of SHE in heavy-ion induced complete-fusion nuclear reactions have been considered. With the DGFRS detection system operating in a real-time mode to search for ER-alpha energy-time-position correlated sequences it became possible to measure 45 new superheavy isotopes in the last ten years. The author considers development of the detection systems, which provide not only event by event data storage, but also on-line (real-time) data processing to be a very positive and fruitful trend in SHE synthesis since these allow radical suppression of beam-associated background. On the other hand, with a rapid development of digital electronics [29], the problem of data compression for real-time processing will be actuated in the nearest future.

Author hopes, that chemical experiments will become more fruitful in the nearest future, especially taking into account progress in minimization of the transport and detection dead time. Only with that improvement it will became possible not only to confirm experiments with electro-magnetic separators, but to discover SHE's.

This paper is supported in part by the RFBR Grant № 11-02-12066.